\documentclass[12pt]{article}
\usepackage{graphicx,graphics}
\usepackage{amssymb}
\usepackage{amsmath}
\usepackage{dcolumn}
\usepackage{bm}
\usepackage{color}
\def\be{\begin{equation}}
\def\ee{\end{equation}}
\newcommand{\bea}{\begin{eqnarray}}
\newcommand{\eea}{\end{eqnarray}}
\newcommand{\nn}{\nonumber}

\def\hbar#1{\backslash\hspace{-2mm}#1}

\def\nn{\nonumber}

\def\2tvec#1#2{
\left(
\begin{array}{c}
#1  \\
#2  \\
\end{array}
\right)}

\def\mat2#1#2#3#4{
\left(
\begin{array}{cc}
#1 & #2 \\
#3 & #4 \\
\end{array}
\right) }

\def\Mat3#1#2#3#4#5#6#7#8#9{
\left(
\begin{array}{ccc}
#1 & #2 & #3 \\
#4 & #5 & #6 \\
#7 & #8 & #9 \\
\end{array}
\right) }

\def\3tvec#1#2#3{
\left(
\begin{array}{c}
#1  \\
#2  \\
#3  \\
\end{array}
\right)}

\def\hbar#1{\backslash\hspace{-2mm}#1}

\def\nn{\nonumber}

\newcommand{\bt}{\begin{itemize}}
\newcommand{\et}{\end{itemize}}
\numberwithin{equation}{section}

\begin{document}

\begin{titlepage}
\begin{flushright}
KIAS-P12066\\
IPPP-12-75\\
DCPT-12-150
\end{flushright}

\begin{center}

\vspace{1cm}
{\large\bf Light Dark Matter Candidate\\ in \\
 $B-L$ Gauged Radiative Inverse Seesaw}
\vspace{1cm}

Yuji Kajiyama,$^{a,}$\footnote{kajiyama-yuuji@akita-pref.ed.jp}
Hiroshi Okada,$^{b,}$\footnote{HOkada@kias.re.kr}
Takashi Toma$^{c,}$\footnote{takashi.toma@durham.ac.uk}
\vspace{5mm}

{\it%
$^{a}${Akita Highschool, Tegata-Nakadai 1, Akita, 010-0851, Japan}\\
$^{b}${School of Physics, KIAS, Seoul 130-722, Korea}\\
$^{c}${Institute for Particle Physics Phenomenology
University of Durham, Durham DH1 3LE, UK}
}
  
  \vspace{8mm}

\abstract{We study a radiative inverse seesaw model with local $B-L$
 symmetry, in which we extend the neutrino mass structure that is generated through a kind of
 inverse seesaw framework to the more generic one than our previous work.
We focus on a real part of bosonic particle as a dark
 matter and investigate the features in ${\cal O}$(1-80) GeV mass range, reported by the experiments such as CoGeNT and XENON (2012). 
 }

\end{center}
\end{titlepage}

\setcounter{footnote}{0}

\section{Introduction}
It is suggested that Dark Matter (DM) exists in our universe from the cosmological
observations such as the rotation curves of the
galaxy~\cite{Begeman:1991iy} and the gravitational
lensing~\cite{Massey:2007wb}, and moreover we know that DM dominates about
23 \% from the CMB observation by WMAP \cite{wmap}. 
However they tell us almost nothing about the scale of DM mass.
In recent years, direct detection experiments of DM such as XENON100 \cite{xenon100},
CRESSTII \cite{cresst}, CoGeNT \cite{cogent} and DAMA \cite{dama} are active to investigate scattering
events with nuclei. XENON100 has not shown a result of DM signal
but shown an upper bound of the scattering cross section with nuclei with the minimal bound around 100 GeV, on the other hand,
CoGeNT, DAMA and CRESSTII have reported the observations which can be interpreted as DM signals that favor a light DM with several GeV mass and rather large cross section.
As far as we consider these experiments, the mass scale of DM should be ${\cal O}$(1-100) GeV.

Here we introduce radiative seesaw model (Ma-Model) \cite{ernest} proposed by Ernest Ma
whose model is known as a TeV scale theory and has an abundant source of DM candidate that includes fermion (that is usually identified to right-handed neutrino; $N^c$)
and boson (that is usually identified to the real/pseudo scalar boson of an additional Higgs with local $SU(2)_L$ symmetry;
$\eta=(\eta^+,(\eta_R+i\eta_I)/\sqrt{2})^T$. 
However, even for both cases, one always has to consider the constraint of the lepton flavor violation (LFV).
The most stringent constraint of LFV comes from the $\mu\rightarrow$ $e\ \gamma$ process mediated by $N^c$ and $\eta^+$
that often leads us not to a scenario with DM of ${\cal O}$(1-10) GeV but of with more than  ${\cal O}$(100-1000) GeV. 
The main reason is that it is difficult to realize the neutrino mixing matrix \cite{mns}, with maintaining the diagonal neutrino Yukawa matrix $(y_\nu)$
that can leads to a light DM scenario. For the other aspect,
a four point coupling of Higgs has to be enough tiny (that requires $\lambda_5\sim 10^{-5}$)
to induce an appropriate neutrino mass scale with ${\cal O}$(1) Yukawa coupling in a TeV scale theory.
It tells us that the mass of $\eta_{R}$ and $\eta_I$ has to be degenerate.
It suggests that an inelastic scattering process of DM via Z-boson can be dominant.
As a result, we have a narrow allowed region; that is a CoGeNT region, in the direct detection experiments.
Such a model is ruled out by the experiments that give an upper bound.
This is because a scenario with DM of ${\cal O}$(1-10) GeV is difficult 
to be realized even in the bosonic case.
One of the simple/straightforward solution is to increase the value of $\lambda_5$ (with diagonal $y_\nu$); that is, the mass of $\eta_{R}$ and $\eta_I$ is hierarchical.
To realize it, we revisit the radiative inverse seesaw model with local gauged $B-L$ symmetry \cite{Okada:2012np}.
This scenario is in fact a quite promising aspect to make a hierarchy between them and to retain the diagonal $y_{\nu}$.
In general, inverse seesaw includes a tiny mass scale of $\mu$ that plays the crucial roles in
explaining the observed neutrino masses and their mixing angles. 
In other words, the smallness of $\lambda_5$ can be replaced by $\mu$, and the non-diagonal structure of $\mu$ allows $y_\nu$ to stay diagonal.
Thus we expect that a light DM scenario works well.

In this paper,
we construct the neutrino mass matrix more generically, (without restricting ourselves to a
concrete structure), maintaining the nonzero matrix
form of $\mu$ and $y_\nu$. As a result, we find that 
a bosonic particle as a DM could be a promising candidate in
the wide range of a few GeV-80 GeV that is in favor of the
the direct detection experiments.
 
This paper is organized as follows. In Section 2, we construct the
radiative inverse seesaw model and its Higgs sector, and
we discuss the constraints from LFV, especially, $\mu\to
e\gamma$ process.
In Section 3, we analyze the DM relic abundance and the direct detection of our DM including all the other constraint.
We summarize and conclude the paper in Section 4.

\section{The Radiative Inverse Seesaw Model}
\subsection{Neutrino Physics and Lepton Flavor Violation}

\begin{table}[thbp]
\centering {\fontsize{10}{12}
\begin{tabular}{||c|c|c|c|c|c|c|c||c|c|c||}
\hline\hline ~~Particle~~ & ~~$Q$~~ & ~~$u^c, d^c $~~ & ~~$L$~~ & ~~$e^c$~~ & ~~ $N^c$~~
 & ~~$S_1$~~ & ~~$S_2$~~ & ~~$\Phi~~ $& ~~$\eta~~ $ & $\chi $\\\hline
$SU(2)_L$&$\bm{2}$ & $\bm{1}$ & $\bm{2}$ & $\bm{1}$ & $\bm{1}$ &
$\bm{1}$ & $\bm{1}$ & $\bm{2}$ & $\bm{2}$ & $\bm{1}$\\\hline
$Y_{B-L}$ & 1/3 & -1/3 & -1 & 1 & 1 & -1/2 & 1/2 & 0 & 0 & -1/2\\\hline
$\mathbb{Z}_2$ & + & + & + & + & $-$ & $-$ & + & + & $-$ & +\\
\hline\hline
\end{tabular}%
} \caption{The particle contents and the charges. 
Notice that a pair of fermions $S_1$ and $S_2$ is required from the
 anomaly cancellation.}
\label{tab:b-l}
\end{table}

We have proposed a radiative inverse seesaw model with $U(1)_{B-L}$ in
Ref.~\cite{Okada:2012np} which is an extended model of a radiative
seesaw model proposed by Ma~\cite{ernest}. 
The particle contents are shown in Tab.~\ref{tab:b-l}. We
add three right-handed neutrinos $N^c$, three pair of fermions $S_1$ and
$S_2$, a $SU(2)_L$ doublet scalar $\eta$ and $B-L$ charged scalar
$\chi$ to the standard model, and $\mathbb{Z}_2$ parity is also imposed
to forbid Dirac neutrino masses between left-handed and right-handed
neutrinos at tree level and stabilize DM candidates. It is assumed that
the doublet scalar $\eta$ does not have vacuum expectation value to
have an exact $\mathbb{Z}_2$ parity at low energy scale.
After the electroweak symmetry breaking~\cite{Okada:2012np}, the mass terms in
neutrino sector become 
\begin{equation}
\mathcal{L}_{m}=y_{\nu}\eta N^cL+MN^cS_1
+\frac{\mu}{2}S_1^2+\frac{\mu'}{2}S_2^2
+{\mu''}S_2\nu_L+\mathrm{h.c.},
\end{equation}
where generation indices are abbreviated, and 
$\mu''\ll\mu,\:\mu'$, thus we neglect the Dirac mass term
$\mu''S_2\nu_L$
\footnote{ Since $\mu''/\mu \sim \mu''/\mu' \sim \langle \Phi \rangle/\langle \chi \rangle \sim 0.05$,
it is reasonable to neglect this term. In a supersymmetric model, however, this term automatically vanishes in appropriate assignments.
We shall publish it elsewhere.}. 
In this situation, the light neutrinos are decoupled with $S_2$ and their masses
are produced radiatively via the interaction with $N^c$ and $S_1$. 
Thus $N^c$ and $S_1$ should be expressed by mass eigenstates. 
The $6\times 6$ mass matrix $\mathcal{M}_{\nu}$ of $(N^c, S_1)$ which is
block-diagonalized by a unitary matrix $\Omega$ is given as 
\begin{equation}
{\cal M}_\nu=\left(
\begin{array}{cc}
   0 & M \\
  M^T & \mu\\
\end{array}
\right),\qquad
\Omega^{T}\mathcal{M}_{\nu}\Omega=\left(
\begin{array}{cc}
M_{+} & 0\\
0 & M_{-}
\end{array}
\right),
\end{equation}
where the matrix $\Omega$ is expressed by the exponential of the matrix
$S$, and expanded up to first order of $S$ as 
\begin{equation}
\Omega=
\frac{1}{\sqrt2} \left(
\begin{array}{cc}
 i(1+S^\dag) & 1-S \\
-i(1-S^\dag) & 1+S \\
\end{array}
\right) + {\cal O}(S^2).
\end{equation}
Under the conditions of assuming $M$ is proportional to unit matrix and
$\mu S^\dagger=-S^T \mu$, $\mu S=-S^*\mu$, $MS^\dagger=-S^TM$,
$MS=-S^*M$ as in Appendix of Ref.~\cite{Catano:2012kw}, the solution for
the matrix $S$ is given as $S=\mu/(4M)$. 
The specific condition that the Dirac mass matrix $M$ is proportional to unit
matrix plays an important role to avoid the constraint of $\mu\to
e\gamma$ as we will see below. 
Then the block-diagonalized matrix $M_+$ and $M_-$ which are completely
diagonalized by unitary matrices $U$ and $V$ are expressed as 
\begin{eqnarray}
M_+=M-\frac{\mu}{2}+{\cal O}(S^2)=U^*m^{\mathrm{diag}}_{+}U^\dagger,
\label{mass-flavor1}\\
M_-=M+\frac{\mu}{2}+{\cal O}(S^2)=V^*m^{\mathrm{diag}}_{-}V^\dagger.
\label{mass-flavor2}
\end{eqnarray}
The flavor eigenstates $N^c$ and $S_1$ are given by the mass
eigenstates $\nu_\pm$ with the masses $m_{i\pm}$ as 
\begin{equation}
\left(
\begin{array}{c}
N^c\\
S_1
\end{array}
\right)=\Omega\left(
\begin{array}{cc}
U & 0\\
0 & V
\end{array}
\right)\left(
\begin{array}{c}
\nu_+\\
\nu_-
\end{array}
\right)\equiv
\left(
\begin{array}{cc}
W_N^+ & W_N^-\\
W_S^+ & W_S^-
\end{array}
\right)\left(
\begin{array}{c}
\nu_+\\
\nu_-
\end{array}
\right).
\end{equation}
The light neutrino mass matrix is given by the unitary matrix as 
\begin{eqnarray}
\left(M_\nu\right)_{\alpha\beta}
&=&\sum_{i=1}^3\frac{\left(y_{\nu}W_N^-\right)_{\alpha i}
\left(y_{\nu}W_N^-\right)_{\beta i}}{(4\pi)^2}
m_{i-}
\left[\frac{m^2_R}{m^2_R-m_{i-}^2}\log\frac{m^2_R}{m_{i-}^2}
-\frac{m^2_I}{m^2_I-m_{i-}^2}\log\frac{m^2_I}{m_{i-}^2}\right]\nonumber\\
&&\!\!\!\!\!\!+\sum_{i=1}^3
\frac{\left(y_{\nu}W_N^+\right)_{\alpha i}
\left(y_{\nu}W_N^+\right)_{\beta i}}{(4\pi)^2}
m_{i+}
\left[\frac{m^2_R}{m^2_R-m_{i+}^2}\log\frac{m^2_R}{m_{i+}^2}
-\frac{m^2_I}{m^2_I-m_{i+}^2}\log\frac{m^2_I}{m_{i+}^2}\right],
\end{eqnarray}
where $m_R$ and $m_I$ are masses of $\eta_{R}$ and $\eta_{I}$.
Since $\mu\ll M$, we can simplify to obtain the approximate light neutrino mass matrix as
\begin{equation}
\left(M_{\nu}\right)_{\alpha\beta}
\!\simeq\!
\frac{\left(y_{\nu} \mu \:\!y^T_{\nu}\right)_{\alpha\beta}}{2(4\pi)^2}
\left[\frac{m_R^2}{M^2}I\left(\frac{m_R^2}{M^2}\right)-
\frac{m_I^2}{M^2}I\left(\frac{m_I^2}{M^2}\right)
\right]\quad
\mathrm{with}\quad
I(x)=\frac{x}{1-x}\left(1+\frac{x\log{x}}{1-x}\right),
\label{eq:neut-mass0}
\end{equation}
where $U^TU=V^TV=1$ is assumed. 
Therefore we can see from Eq.~(\ref{eq:neut-mass0}) that the flavor
structure of the neutrino mixing matrix is determined by both of the structure
of $\mu$ and the neutrino Yukawa matrix $y_{\nu}$. 

The most stringent constraint of LFV comes
from $\mu\to e\gamma$ process.
The experimental upper bounds of the branching ratio is
$\mathrm{Br}\left(\mu\to e\gamma\right)\leq2.4\times
10^{-12}$~\cite{Adam:2011ch}. Due to the specific assumption for the
matrix $M$ above, the branching ratio of the process in our
model is calculated as
\begin{equation}
{\rm Br}(\mu\to e\gamma)
=\frac{3\alpha_{\mathrm{em}}}{64\pi(G_{F}M^{2}_{\eta})^{2}}
\left|\left(y_{\nu}y_{\nu}^\dag\right)_{\mu e}\right|^2
F_2^2\left(\frac{M^2}{M_{\eta}^2}\right),
\label{LFV1}
\end{equation}
with
\begin{equation}
F_2(x)=\frac{1-6x+3x^2+2x^3-6x^2\log{x}}{6(1-x)^4},
\end{equation}
where $M_\eta$ is $\eta^+$ mass. 
We can see from the formula that if the Yukawa matrix $y_\nu$ is diagonal, all
LFV processes vanish even though $y_{\nu}\sim1$. 
In the radiative seesaw model~\cite{ernest}, the Yukawa matrix must have a structure
in order to derive a non trivial mixing matrix of neutrinos. On the other
hand, it is possible that the Majorana mass term $\mu$ has a flavor structure instead of
$y_{\nu}$ in the radiative inverse seesaw model. Thus deriving neutrino mixing matrix and
the diagonal matrix of $y_\nu$ can be consistent with each other even if
$y_\nu$ is diagonal\footnote{With the non-diagonal structure of $y_\nu$, the LFV constraint can be satisfied under the condition that $M_\eta\simeq$ 
${\cal O}$(500) GeV for $y_\nu\le$0.1. In this case, however, 
our DM ($\eta_R$) mass cannot be of ${\cal O}$(1-10) GeV because the mass difference between $m_R$ and $M_\eta$ is proportional to $v$.}. 

\subsection{Scalar Potential}
The scalar Higgs potential of this model is given by~\cite{Okada:2012np}
\begin{eqnarray}
V \!\!\!&=&\!\!\!
 m_1^{2} \Phi^\dagger \Phi + m_2^{2} \eta^\dagger \eta  + m_3^{2} \chi^\dagger \chi +
\lambda_1 (\Phi^\dagger \Phi)^{2} + \lambda_2 
(\eta^\dagger \eta)^{2} + \lambda_3 (\Phi^\dagger \Phi)(\eta^\dagger \eta) 
+ \lambda_4 (\Phi^\dagger \eta)(\eta^\dagger \Phi)
\nonumber \\ &&\!\!\!
+
\lambda_5 [(\Phi^\dagger \eta)^{2} + \mathrm{h.c.}]+
\lambda_6 (\chi^\dagger \chi)^{2} + \lambda_7  (\chi^\dagger \chi)
(\Phi^\dagger \Phi)  + \lambda_8  (\chi^\dagger \chi) (\eta^\dagger \eta),
\end{eqnarray}
where $\lambda_5$ has been chosen real without any loss of
generality. $\lambda_1$, $\lambda_2$ and $\lambda_6$ have to be positive
to stabilize the Higgs potential.  
After the symmetry breaking $\Phi^0=\left(v+\phi^0\right)/\sqrt{2}$ and
$\chi=\left(v'+\chi^0\right)/\sqrt{2}$, the gauge eigenstates $\phi^0$
and $\chi^0$ mix and are rewritten in terms of the mass
eigenstates of the SM-like Higgs $h$ and an extra heavy Higgs $H$ as
\begin{eqnarray}
\phi^0 &=& h\cos\alpha + H\sin\alpha, \nn\\
\chi^0 &=&-h\sin\alpha + H\cos\alpha.
\label{eq:mass_weak}
\end{eqnarray}
Three-point couplings of $h\eta_R\eta_R$ and $H\eta_R\eta_R$ are important
to investigate DM analysis since we identify $\eta_R$ which is real
part of $\eta^0$ is DM candidate. We define $h\eta_R\eta_R$ coupling as
$\lambda_h$ and $H\eta_R\eta_R$ coupling as $\lambda_H$ and these are
written as
\footnote{The Higgs analysis in the typical inverse seesaw model has
been done in the Ref. \cite{Bandyopadhyay:2012px, Gogoladze:2012jp},
in which the detectability of the recent experiments ATLAS and CMS is discussed.} 
\begin{eqnarray}
2\lambda_h=\left(\lambda_3+2\lambda_4+2\lambda_5\right)v\cos\alpha-\lambda_8v'\sin\alpha,\\
2\lambda_H=\left(\lambda_3+2\lambda_4+2\lambda_5\right)v\sin\alpha+\lambda_8v'\cos\alpha.
\end{eqnarray}

\section{Dark Matter}
\subsection{DM Relic Density}
There are several DM candidates such as $\nu_{1-}$, $\eta_R$ and $\eta_I$
in the model where $\eta_R$ and $\eta_I$ are real and imaginary part of
the neutral component of $\eta$. The sign of $\lambda_5$ determines that
either $\eta_R$ or $\eta_I$ becomes DM candidate. 
We identify that DM is $\eta_R$ here whose mass is less than $W$ boson
mass since the relic density is severely reduced due to the other
annihilation processes if the DM mass is larger than that.
The fermionic DM $\nu_{1-}$ with
degenerated $\nu_{1+}$ have been investigated in Ref.~\cite{Okada:2012np}. In a
similar radiative seesaw model~\cite{ernest}, the mass
difference between $\eta_R$ and $\eta_I$, which is proportional to
$\lambda_5 v^2$, must be small enough because it is
correlated with generating tiny neutrino masses. 
We do not need such a
small mass difference of $\eta_R$ and $\eta_I$ in the radiative inverse
seesaw model to generate the proper neutrino masses because $N^c$ and
$S_1$ are degenerated instead of $\eta_R$ and $\eta_I$. 

\begin{figure}[t]
\begin{center}
\includegraphics[scale=1]{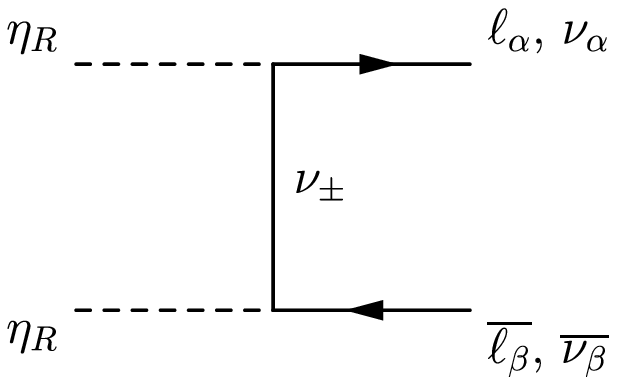}
\qquad
\includegraphics[scale=1]{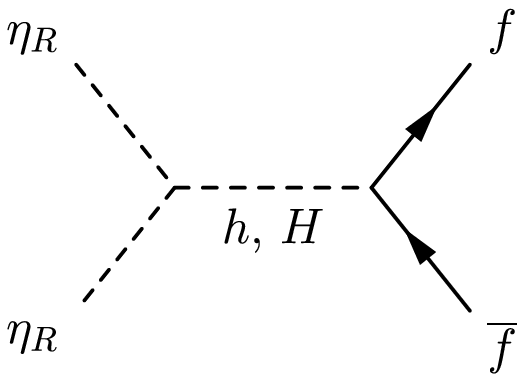}
\caption{The annihilation processes of DM $\eta_R$.}
\label{fig:ann}
\end{center}
\end{figure}

We assume that the DM mass is less than $W$ boson mass, otherwise the
annihilation cross section is too large to satisfy the DM relic
abundance unless a TeV scale DM is taken into account. 
In our case there are two annihilation processes
via $\eta$ and Higgs exchange as shown in Fig.~\ref{fig:ann}, and 
the annihilation cross sections for each process are calculated as
\begin{eqnarray}
\sigma_1v_{\rm rel}
\!\!\!&\simeq&\!\!\!
\frac{\mathrm{Tr}\left(y_\nu y_\nu^\dag y_\nu y_\nu^\dag\right)m_{DM}^4}
{2\pi\left(m^2_{DM}+M^2\right)^3}v_{\mathrm{rel}}^2,
\label{eq:ann1}
\\
\sigma_2v_{\rm rel}
\!\!\!&\simeq&\!\!\!
\sum_{f}\frac{c_fy^2_f}{2\pi}
\left|\frac{\lambda_h\cos\alpha}{4m_{DM}^2-m_{h}^2+im_{h}\Gamma_h}
+\frac{\lambda_H\sin\alpha}{4m_{DM}^2-m_H^2+im_{H}\Gamma_H}\right|^2\left(1-\frac{m_{f}^2}{m_{DM}^2}\right)^{3/2},
\label{eq:ann2}
\end{eqnarray}
where $m_{DM}=m_R$ is DM mass,
$y_f$ is Yukawa coupling of SM matter particle, and 
the color factor $c_f$ is $3$ for quarks and $1$ for leptons, and
the interference terms among the processes are neglected. 
Notice that $\lambda_h$ and $\lambda_H$ have a mass dimension.
The total cross section is $\sigma
v_{\mathrm{rel}}=\sigma_1v_{\mathrm{rel}}+\sigma_2v_{\mathrm{rel}}$. 
The SM-like Higgs mass and total decay width are fixed to
$m_h=125~\mathrm{GeV}$ and $\Gamma_h=4.1\times10^{-3}~\mathrm{GeV}$~\cite{higgsdecay}. 
The decay width of the heavy Higgs $\Gamma_H$ is expressed as 
\begin{equation}
\Gamma_H(H\to
 f\overline{f})=\sum_{f}\frac{c_fy_{f}^2\cos^2\alpha}{16\pi}m_H\left(1-\frac{4m_{f}^2}{m_H^2}\right)^{3/2}.
\end{equation}
The contribution of the process $H$ $\to$ 2DM is also added
to the decay width when the relation $m_H\geq2m_{DM}$ is satisfied.
We can take $y_{\nu}\sim1$, however the Dirac neutrino mass $M$ should be TeV
scale which is correlated with $U(1)_{B-L}$ breaking scale. As a result
$\sigma_1v_{\mathrm{rel}}$ becomes much smaller than the proper value to
reduce to the observed DM relic density. Thus the Higgs exchange process
is only taken into account for the DM annihilation\footnote{If one would
like to consider the constraint of the anti-proton no excess
reported by PAMELA, the existence of s-wave is not favor in the NFW
profile \cite{Lavalle:2010yw, Okada:2012cc}. But one can easily escape
such a constraint to adopt another profile. }. 


\begin{figure}[t]
\begin{center}
\includegraphics[scale=0.65]{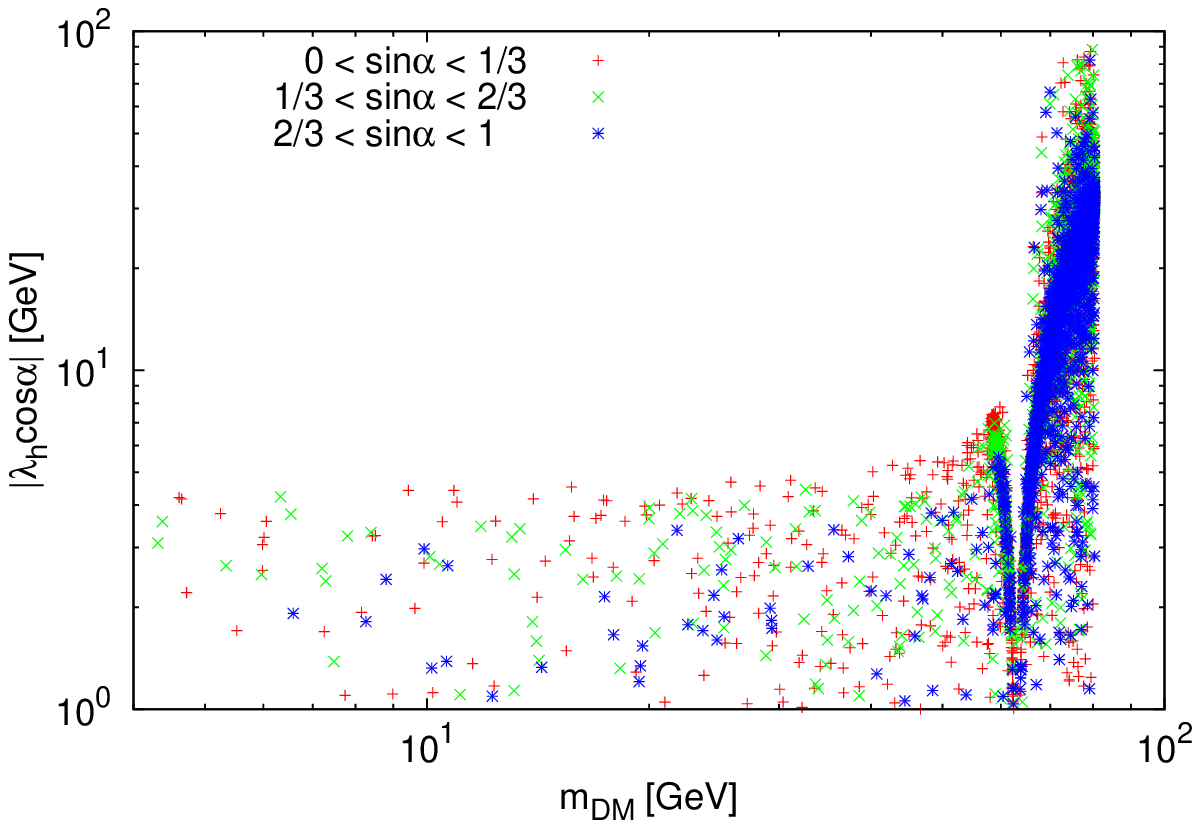}
\qquad
\includegraphics[scale=0.65]{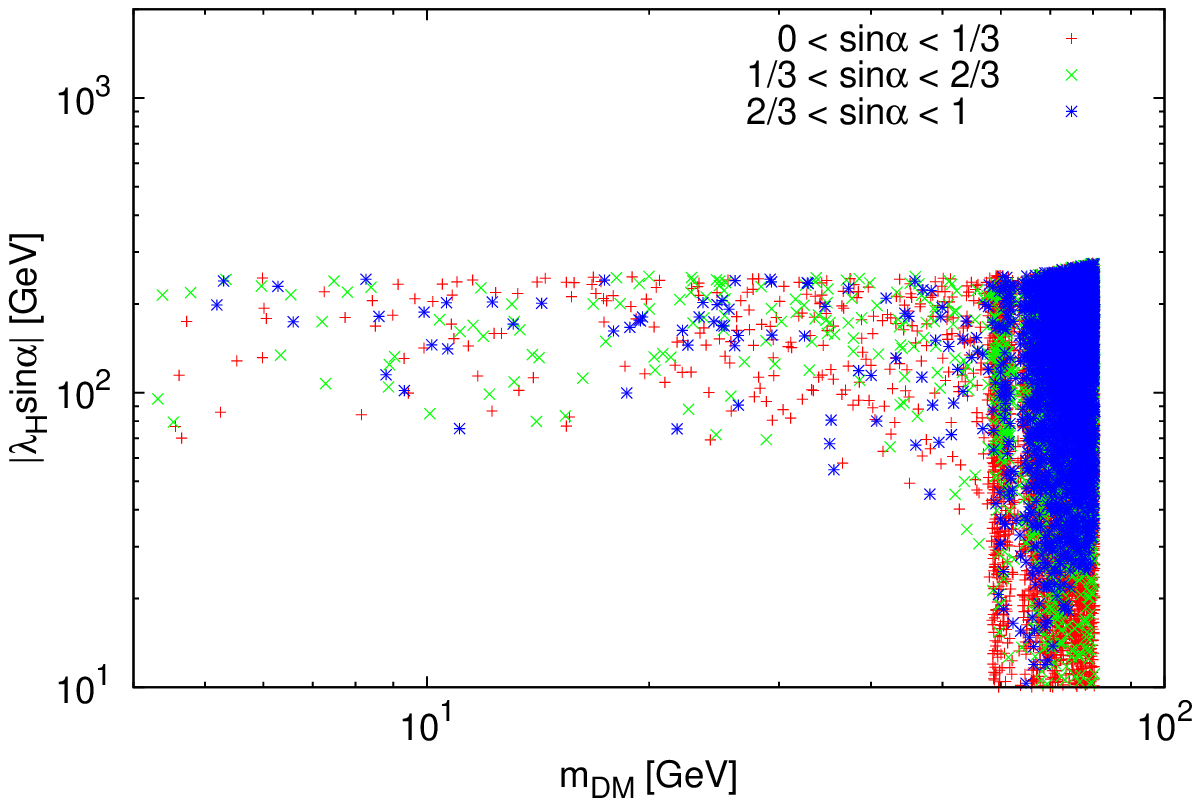}
\caption{The parameter space which satisfies
 $\sigma v_{\mathrm{rel}}\approx3.0\times10^{-26}~\mathrm{cm^3/s}$ and
 ${\rm BR_{inv}}<0.4$ in the $(m_{DM}$-$\lambda_h\cos\alpha)$ and
 $(m_{DM}$-$\lambda_H\sin\alpha)$ plane. 
The VEV of $\chi$ is fixed to $v'=5~\mathrm{TeV}$.}
\label{fig:relic}
\end{center}
\end{figure}

We run the parameters $\lambda_i$, $\alpha$, $m_H$, $m_{DM}$ to explore
a parameter space which satisfies the certain DM relic density observed
by WMAP, which corresponds to $\sigma v_{\mathrm{rel}}\approx3.0\times10^{-26}~\mathrm{cm^3/s}$ 
of the annihilation cross section of DM. 
The allowed parameter region in the DM mass and coupling plane are shown
in Fig.~\ref{fig:relic}.
The process through the SM-like Higgs $h$ is important to satisfy the
constraint since the Yukawa coupling $y_f$ is not so large. Thus we need
a resonance at $2m_{DM}\approx m_h$. 
We can see from Fig.~\ref{fig:relic} that the annihilation cross section
is almost determined by the interaction with the SM like Higgs and the
contribution via the extra Higgs $H$ is small enough. 

\subsection{Invisible Decay of SM-like Higgs}
Recently LHC reported that an invisible decay of SM-like Higgs severely
restricted. The invisible branching ratio $\rm{BR_{inv}}$ is excluded to
the region $0.4 \lesssim {\rm BR_{inv}}$~{\cite{Giardino:2012dp}. 
In our case, SM-like Higgs $h$ can decay into 2DM if the
mass relation $m_{DM}<m_h/2$ is satisfied, and this mode is invisible. 
Our invisible decay width is given by
\bea
\Gamma(h\rightarrow2\mathrm{DM})\simeq\frac{\lambda_h^2}{16\pi m_h}\sqrt{1-\frac{4m_{DM}^2}{m_h^2}}.
\eea
This implies that the invisible decay almost directly constrains the
coupling $\lambda_h$ since the mass of SM-like Higgs is fixed to
$125~\mathrm{GeV}$. 
One can see from the left hand side of Fig.~\ref{fig:relic} that the
coupling is roughly constrained to
$\lambda_h\cos{\alpha}\lesssim10~\mathrm{GeV}$ when
$m_{DM}\lesssim m_h/2$. 
On the other hand, the large value of the coupling $\lambda_H\sin\alpha$ is only
allowed as we see from the right hand side. 


\subsection{Direct Detection}
The DM becomes a Higgs portal DM from the analysis of the DM relic
density. We investigate the detection property by direct detection
experiments of DM. The elastic cross section with a proton occurs via Higgs
exchange and is calculated as 
\begin{equation}
\sigma_p=\frac{\mu_{\mathrm{DM}}^2}{\pi}\frac{m_p^2}{m_{DM}^2v^2}
\left(\frac{\lambda_h\cos\alpha}{m_h^2}+\frac{\lambda_H\sin\alpha}{m_H^2}\right)^2
\left(\sum_qf_q^p\right)^2,
\end{equation}
where $\mu_{\mathrm{DM}}=\left(m_p^{-1}+m_R^{-1}\right)^{-1}$ is
proton-DM reduced mass and the parameters $f_q^p$ are determined from
the pion-nucleon sigma term $\sigma_{\pi N}$ as
\begin{equation}
f_u^p=0.023,\quad
f_d^p=0.032,\quad
f_s^p=0.020
\end{equation}
for light quarks and $f_Q^p=2/27\left(1-\sum_{q\leq3}f_q^p\right)$ for
heavy quarks where $Q=c,b,t$~\cite{pdg}.

We compare the numerical result which satisfies the DM relic abundance
with several direct detection experiments as shown in
Fig.~\ref{fig:sigma}. 
We can see from the figure that the Higgs portal DM has a large elastic
cross section which can be consistent with  CRESSTII~\cite{cresst}, CoGeNT~\cite{cogent}, and DAMA~\cite{dama}. 
The XENON100 curve implies the upper bound for the elastic cross
section. Although the bound is quite strong, a small allowed parameter space
exists, especially nearby $m_{DM}\sim m_h/2$. This is the resonance
point of SM-like Higgs exchange process. 
On the other hand, all the experiments can not be consistent with each other unless
taking into account a specific property of DM, thus DM nature is
not clear experimentally yet. 
Further DM properties will be revealed by future direct detection
experiments such as XENON1T~\cite{xenon1t}. 

\begin{figure}[t]
\begin{center}
\includegraphics[scale=0.65]{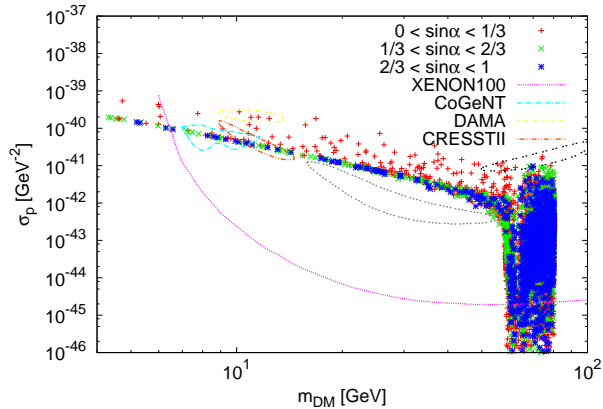}
\caption{The comparison of the elastic cross section with the direct
 detection experiments. XENON100, CoGeNT, DAMA and CRESSTII results are shown here.}
\label{fig:sigma}
\end{center}
\end{figure}
\label{secdirect}

\section{Conclusions and Discussions}
We constructed the neutrino mass matrix more generically, (without restricting ourselves to a
concrete structure), maintaining the non-diagonal matrix
form of $\mu$ with diagonal $y_\nu$ in the frame of a radiative inverse seesaw model with local $B-L$ symmetry. 
We showed that the mass of $\eta_{R}$ can be possible to be smaller than that of $\eta_I$,
since $\lambda_5$ can be taken to be of ${\cal O}$(1). Moreover the
severe constraint of the LFV such as $\mu\rightarrow e\gamma$ process
can be relaxed because of the diagonal $y_\nu$.
As a result, we found a real part of bosonic particle $\eta_R$ as a DM
can be a promising candidate in the wide allowed mass range of ${\cal
O}$(1-80) GeV that is in favor of the the direct detection experiments
such as CRESSTII, CoGeNT and DAMA. 

In the typical inverse seesaw model \cite{non-susy-inverse-pheno2}, the
Yukawa coupling $y_\nu$ would be constrained by the lepton universality that tells us $y_\nu<$0.1 \cite{delAguila:2008pw}.
It generally occurs as far as a Dirac mass term of $N^c$ exists. However we can easily evade such a constraint, 
since we have no Dirac mass term. On the other hand, the typical inverse seesaw can explain the muon anomalous magnetic moment well 
through the diagram via the charged gauge boson and (six) additional Majorana particles \cite{Abdallah:2011ew},
due to the mixing between active neutrinos and the other ones.
In this aspect, our model is not in favor of the muon anomalous magnetic moment.

\section*{Acknowledgments}
H.O. thanks to Prof. Eung-Jin Chun, Dr. Priyotosh Bandyopadhyay, and
Dr. Jong-Chul Park, for fruitful discussion.
Y.K. thanks Korea Institute for Advanced Study for the travel support and local hospitality
during some parts of this work.

\end{document}